\documentclass{article}
\usepackage{locata,amsmath,amsfonts,amssymb, graphicx,url,times}

\DeclareMathOperator*{\argmax}{arg\,max}

\title{Evaluating MCC-PHAT for the LOCATA Challenge - Task 1 and Task 3}

\name{Shoufeng Lin}
\address{Curtin University\\
School of Electrical Engineering, Computing and Mathematical Sciences \\
Bentley, Western Australia \\
shoufeng.lin@postgrad.curtin.edu.au}

\begin{document}

\ninept
\maketitle

\begin{sloppy}

\begin{abstract}
This report presents test results for the \mbox{LOCATA} challenge \cite{lollmann2018locata} using the recently developed MCC-PHAT (multichannel cross correlation - phase transform) sound source localization method. 
The specific tasks addressed are respectively the localization of a single static and a single moving speakers using sound recordings of a variety of static microphone arrays. 
The test results are compared with those of the MUSIC (multiple signal classification) method. 
The optimal subpattern assignment (OSPA) metric is used for quantitative performance evaluation. 
In most cases, the MCC-PHAT method demonstrates more reliable and accurate location estimates, in comparison with those of the MUSIC method.
\end{abstract}

\begin{keywords}
localization, MCC-PHAT, moving speaker, MUSIC, microphone array processing
\end{keywords}

\section{Introduction}
\label{sec:intro}

There is a significant body of literature for the problem of sound source localization, including acoustic speaker localization in particular. 
Obtaining accurate location estimates enables further signal processing, e.g. speaker tracking, speech separation via beamforming, as well as speech enhancement and dereverberation. It also has wide practical applications, e.g. automatic camera steering in smart environments, hearing aids and smart home assistance, as well as virtual reality synthesis. 

In a recent localization paper \cite{lin2018joint, lin2018reverb}, the author investigated the reverberation-robust localization approach of using redundant information of multiple microphone pairs, and proposed the Onset-MCCC and MCC-PHAT methods. The performance of the proposed methods has been evaluated for static and moving speakers in various reverberant scenarios, using sound recordings from a uniform circular array (UCA). Compared with some state-of-the-art location estimators, e.g. the EB-ESPRIT \cite{teutsch2006acoustic}, TF-CHB \cite{torres2012robust} and Neuro-Fuzzy \cite{plinge2010robust}, the proposed methods demonstrate encouraging localization capabilities.

In this report, the sound corpus from LOCATA \cite{lollmann2018locata} is used for evaluating the performance of the MCC-PHAT method. 
Sound recordings from different microphone arrays are used. Tasks of single static speaker and single moving speaker are addressed.
The OSPA metric \cite{schuhmacher2008consistent} was used in \cite{lin2018reverb} for evaluating localization performance in complicated scenarios, e.g. time-varying number of active speakers. In this paper, since the number of speakers is constant (only one speaker), the OSPA metric can be simplified to the root mean square (RMS).

\section{Problem Formulation and Localization Algorithms}
\label{sec:mccphat}

\subsection{Signal Model}

Assuming far-field planar wave speaker signals, the sound recording from an arbitrary microphone $i$ ($i=1,\cdots,I_M$, $I_M$ is the total number of microphones) can be a mixture of speaker sound signals and noises. For a series of discrete time samples, 
\begin{subequations}
\begin{align}
x_i (n/f_s) & = y_q(n/f_s - \tau_{qi}) + n_i(n/f_s) , 
\intertext{or in the short time Fourier transform (STFT) domain}
X_i (\omega_n) & = e^{-\jmath \omega_n \tau_{qi}} \cdot Y_q(\omega_n) + N_i(\omega_n) , 
\end{align}
\end{subequations}
where $\jmath=\sqrt{-1}$, $f_s$ is the sampling frequency, $n =0, 1, \cdots, N-1$, $N$ is the number of samples for the STFT, $\tau_{qi}$ is the time delay from speaker to microphone $i$, $\omega_n = 2 \pi f_s n / N $ is the angular frequency. $y_q(n)$ and $Y_q(\omega_n)$ are the speaker sound signal, and $n_i(t)$ and $N_i(\omega_n)$ are the additive noise at microphone $i$. For notational simplicity, the frame index is suppressed in the STFT expression.

\subsection{Localization Algorithms}

Many algorithms rely on the signal covariance matrix, i.e.
\begin{equation} \label{eq:covarianceMat2}
\mathbf{R} \triangleq \mathbb{E} [ \mathbf{X} \mathbf{X}^H ] \approx \big< \mathbf{X}(\omega_n) \mathbf{X}^H(\omega_n) \big> ,  
\end{equation}
where $\mathbb{E}[\cdot]$ denotes mathematical expectation, $\big< \cdot \big>$ denotes time average, the ergodicity assumption is used for the approximation, and 
\begin{equation}
\mathbf{X} = [X_1, X_2, \cdots, X_{I_M}]^T . 
\end{equation}

Assuming wide sense stationarity in short time frames, and that the number of active sources is smaller than $I_M$, thus the eigenvalues and eigenvectors of the covariance matrix are obtained via eigendecomposition, i.e. 
\begin{equation}
\mathbf{R} ~ \mathbf{V} = \mathbf{V} ~ \mathbf{\Lambda} , 
\end{equation}
where $\mathbf{\Lambda}$ is a diagonal matrix containing the eigenvalues of $\mathbf{R}$ in descending order, columns of matrix $\mathbf{V}$ are the corresponding eigenvectors of $\mathbf{R}$. 

In particular, 
\begin{equation}
\label{eq:subspaces}
\mathbf{V} \triangleq [\mathbf{E}_S | \mathbf{E}_N],
\end{equation} 
where $\mathbf{E}_S$ is an $I_M \times \hat{Q}$ matrix ($\hat{Q}$ is the estimated number of speakers), while $\mathbf{E}_N$ is an $I_M \times \hat{N}$ matrix ($\hat{N} = I_M - \hat{Q}$). Column vectors of the $\mathbf{E}_S$ and $\mathbf{E}_N$ correspond to the descending order of eigenvalues, span the signal subspace and noise subspace respectively, and are orthonormal.

\subsubsection{MUSIC}

The classical MUSIC method \cite{schmidt1986multiple} formulates the localization function,
\begin{equation}
\varepsilon_0^{\mathrm{music}}(\theta, \omega_n) = \frac{1}{ \mathbf{d}^{H}(\theta, \omega_n) ~ \mathbf{E}_N \mathbf{E}_N^{H} ~ \mathbf{d}(\theta, \omega_n) } . 
\end{equation}
where $\theta$ is the direction of arrival (DOA), and the steering vector is 
\begin{equation}
\mathbf{d}(\theta, \omega_n) = [e^{-\jmath \omega_n \tau_{1}(\theta)}, \cdots, e^{-\jmath \omega_n \tau_{I_M}(\theta)} ] . 
\end{equation}

For wideband applications, the localization function can be expressed as:
\begin{equation}
\varepsilon^{\mathrm{music}}(\theta) = \sum_{\omega_n} \varepsilon_0^{\mathrm{music}} (\theta, \omega_n) . 
\end{equation}
The implementation of the MUSIC method as provided by the LOCATA challenge is used as the reference method.

\subsubsection{MCC-PHAT \cite{lin2018reverb}}


%
The standard cross-correlation function between two signals is
\begin{equation}
R_{x_i x_j}(\tau) \triangleq \mathbb{E} [ x_i(t) x_j(t-\tau) ] ,
\end{equation}
where the goal for time delay estimation is the find the time delay $\tau$ that corresponds to the maximum cross-correlation output. 

The classical generalized cross-correlation (GCC) method uses the cross-power spectral density (CSD) function. Compared with the standard cross-correlation function, it improves the performance of the time delay estimation by pre-filtering signals prior to cross correlation \cite{knapp1976generalized, carter1987coherence}. 

Assuming that the signal and noise power spectral density  (PSD) do not vary significantly over frequency ranges, and the observation time $N / f_s$ is much longer than the possible time delays, the CSD between observed signals is \cite{knapp1976generalized}
\begin{equation} \label{eq:Gx1x2}
S_{x_i x_j}(\omega_n) \approx \frac{N}{f_s} \cdot \mathbb{E} [ X_i(\omega_n) \cdot X_j^{\star}(\omega_n) ] .
\end{equation}

The CSD between filtered outputs is 
\begin{equation} \label{eq:GCC}
G_{ij}^{\mathrm{gcc}}(\omega_n) = 
{\Psi_{ij}(\omega_n)} \cdot {S_{x_i x_j}(\omega_n)} ,
\end{equation}
where the prefilter response is $\Psi_{ij}(\omega_n) = 1/ |S_{x_i x_j}(\omega_n)|$ for PHAT. 

Using the relationship between cross-correlation and CSD
\begin{equation}
R_{x_i x_j}(\tau) = \int _{-\infty} ^{+\infty} S_{x_i x_j}(\omega) \cdot e^{\jmath \omega \tau} ~d\omega ,
\end{equation}
the TDOA estimation between two microphones is thus
\begin{equation} \label{eq:tdePhat}
\hat{\tau}^{\mathrm{gcc}}_{ij} = \argmax _ {\tau} ~ \tilde{R}_{ij}^{\mathrm{gcc}} (\tau) ,
\end{equation}
where for discrete time signals, 
\begin{equation} \label{eq:ximccphat}
\tilde{R}_{ij}^{\mathrm{gcc}} (\tau) = 
\sum _{n=0} ^{N-1} G_{ij}^{\mathrm{gcc}}(\omega_n) \cdot e^{\jmath \omega_n \tau} . 
\end{equation}
Here $\hat{\tau}^{\mathrm{gcc}}_{ij}$ is the estimated time delay between the $i$-th and $j$-th microphones, $[\cdot]^{\star}$ the complex conjugate operation.

The overall localization function of the MCC-PHAT is
\begin{equation} \label{eq:mcc-phat}
\varepsilon^{\mathrm{mcc-phat}} (\theta) \triangleq  \prod \limits_{(i,j) \in \mathbf{P} }  \Bigl\lfloor \tilde{R}_{ij}^{\mathrm{gcc-phat}} (\tau(\theta))  \Bigr\rfloor ,
\end{equation}
where $\tau$ is a function of $\theta$, and the set of microphone pairs $\mathbf{P}$ is given in (\ref{eq:micPair}), for avoiding spatial alias.
\begin{equation}
\label{eq:micPair}
\mathbf{P} = \{ (i,j) \; | \| \vec{m}_i - \vec{m}_j \| < v/f_{max} ) ;  \; i<j \} ,
\end{equation}
where $v$ is the velocity of sound, $\vec{m}_i$ is the location of microphone $i$, and $f_{max}$ is the maximum signal frequency considered.

\section{Numerical Results}
\label{sec:results}

This section presents the localization results of the MCC-PHAT method using sound recordings from various microphone arrays, i.e. the ``benchmark2'', ``dicit'' and ``dummy''. Results using the ``Eigenmike'' are not provided here, as it has many microphones, which suits particular localization algorithms and may otherwise require very long computation time for other methods.

\subsection{Single Static Speaker - ``Task 1''}
This subsection demonstrate the performance of the MCC-PHAT method for localization of a single static speaker, which is referred to as ``Task 1'' in the LOCATA challenge. Due to availability of recordings, test results for the benchmark2 and dicit microphone arrays using ``recording1'', while results for the dummy microphone array using ``recording4'' are plotted as follows.

\subsubsection{Benchmark2}

Fig.~\ref{fig:Task1Benchmark} shows the test results using the benchmark2 microphone array. 
\begin{figure}[h!]
  \centering
  \centerline{\includegraphics[width=\columnwidth]{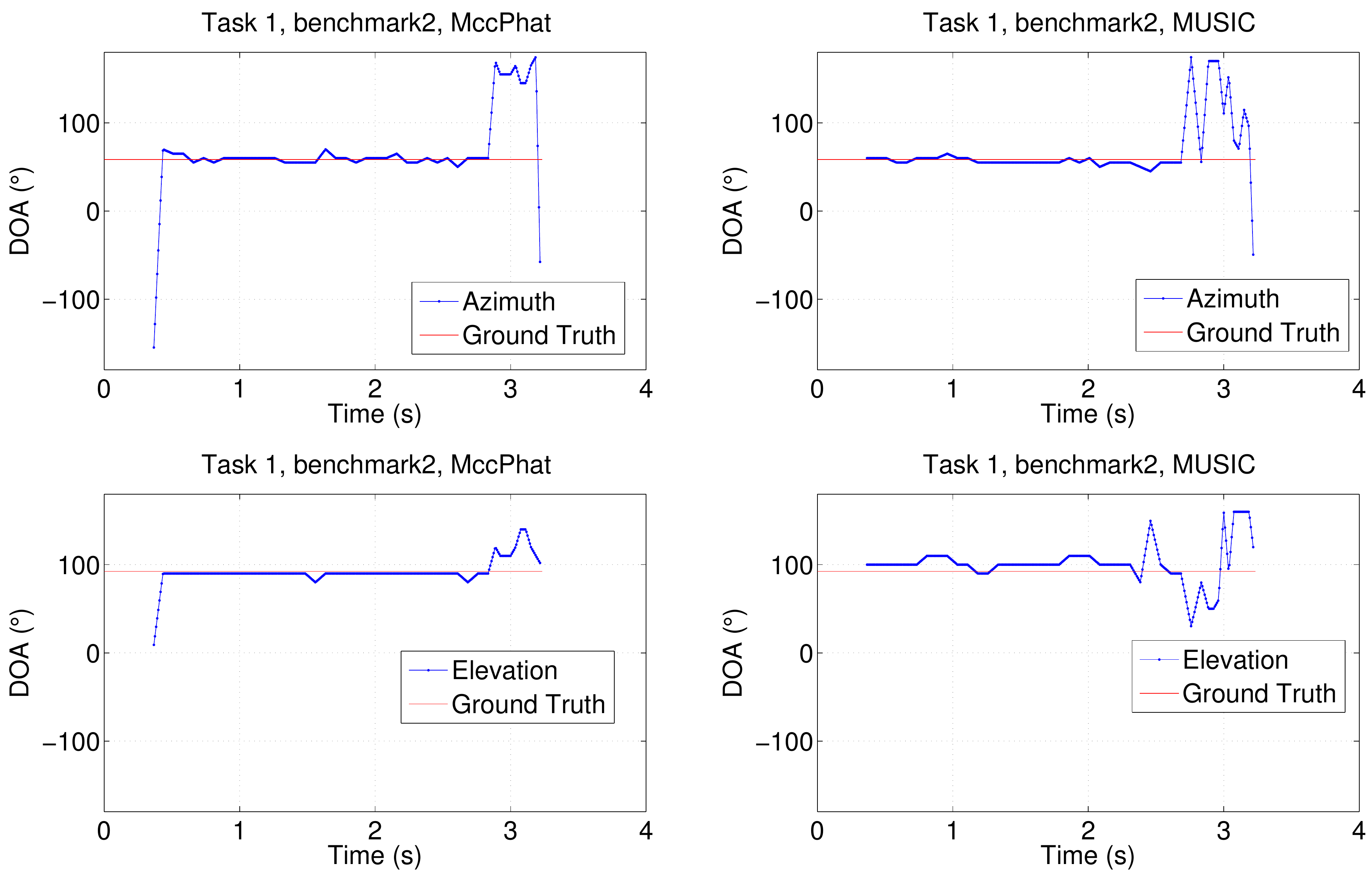}}
  \caption{Test results using benchmark2 microphone array.}
  \label{fig:Task1Benchmark}
\end{figure}

\subsubsection{DICIT}

Fig.~\ref{fig:Task1dicit} shows the test results using the dicit microphone array. 
\begin{figure}[h!]
  \centering
  \centerline{\includegraphics[width=\columnwidth]{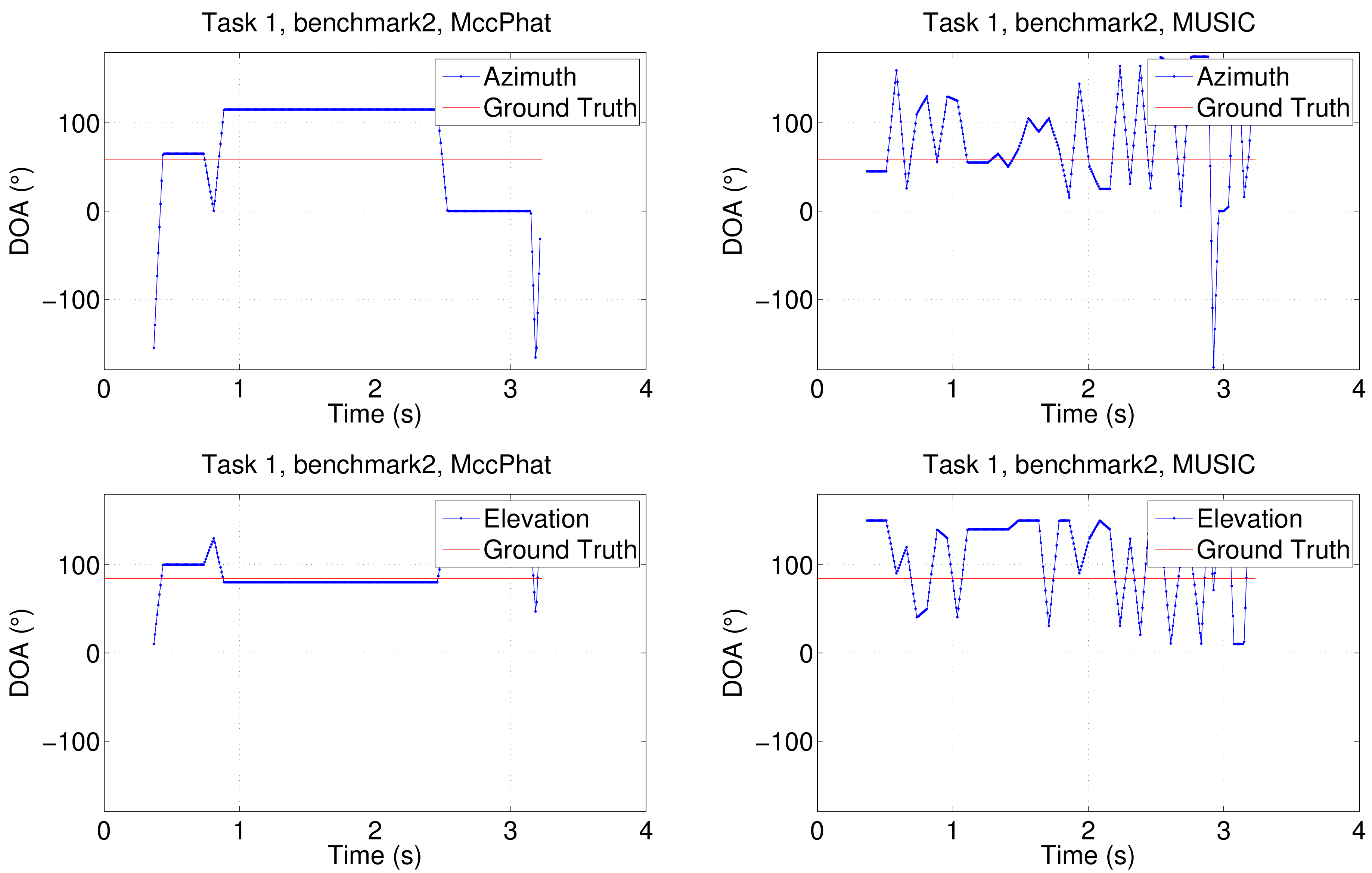}}
  \caption{Test results using dicit microphone array.}
  \label{fig:Task1dicit}
\end{figure}

\subsubsection{Dummy}

Fig.~\ref{fig:Task1dummy} shows the test results using the dummy microphone array. 
\begin{figure}[h!]
  \centering
  \centerline{\includegraphics[width=\columnwidth]{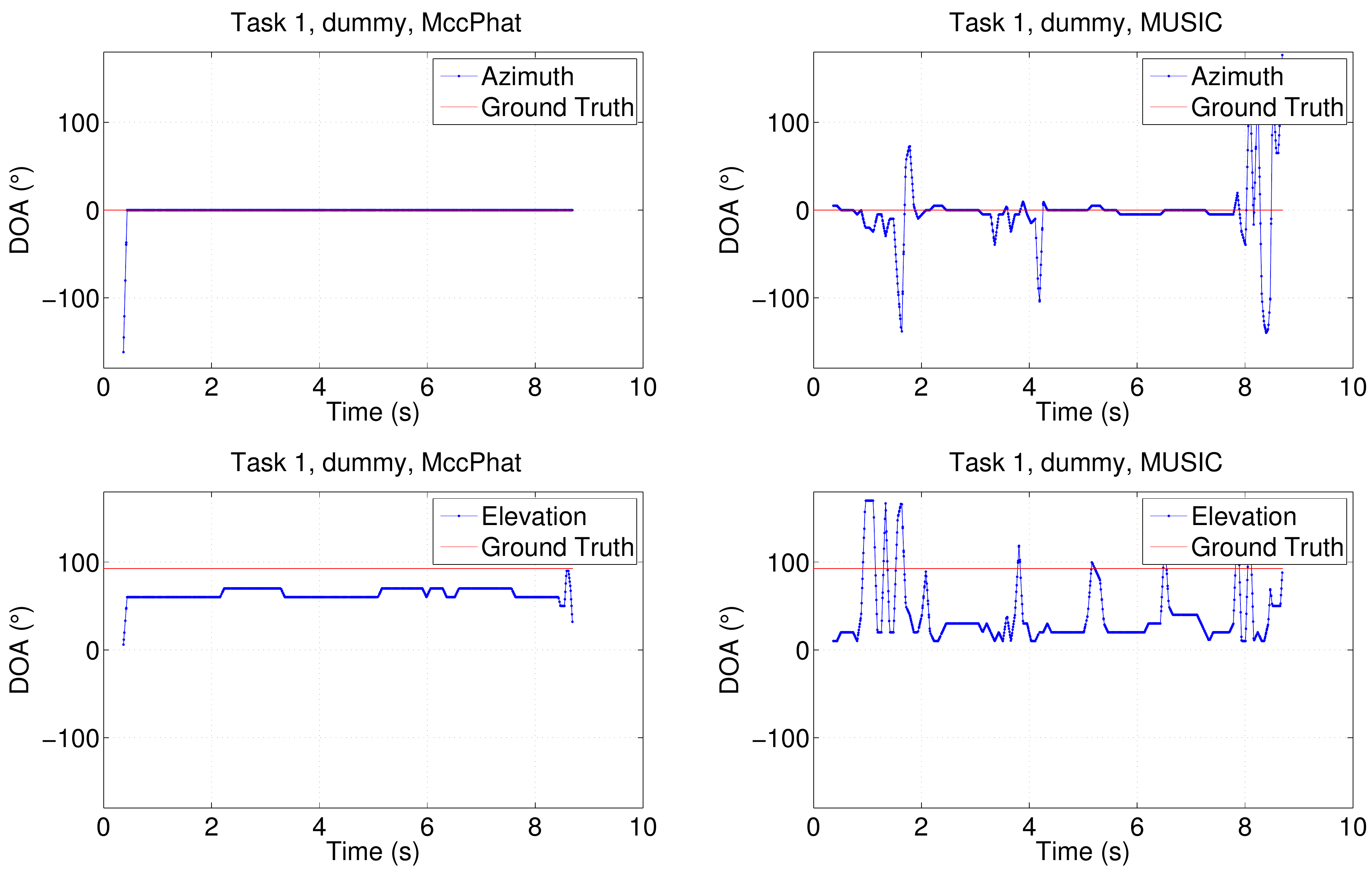}}
  \caption{Test results using dummy microphone array.}
  \label{fig:Task1dummy}
\end{figure}

Fig.~\ref{fig:Task1Benchmark} to~\ref{fig:Task1dummy} indicate that the MCC-PHAT method seems more reliable than the MUSIC method, for the localization of a static speaker. Quantitative analysis is given in Section~\ref{sec:ospa}.


\subsection{Single Moving Speaker - ``Task 3''}
This subsection demonstrate the performance of the MCC-PHAT method for localization of a single moving speaker, which is referred to as ``Task 3'' in the LOCATA challenge. Test results using ``recording2'' are plotted as follows.

\subsubsection{Benchmark2}

Fig.~\ref{fig:Task3benchmark} shows the test results using the benchmark2 microphone array. 
\begin{figure}[!h]
  \centering
  \centerline{\includegraphics[width=\columnwidth]{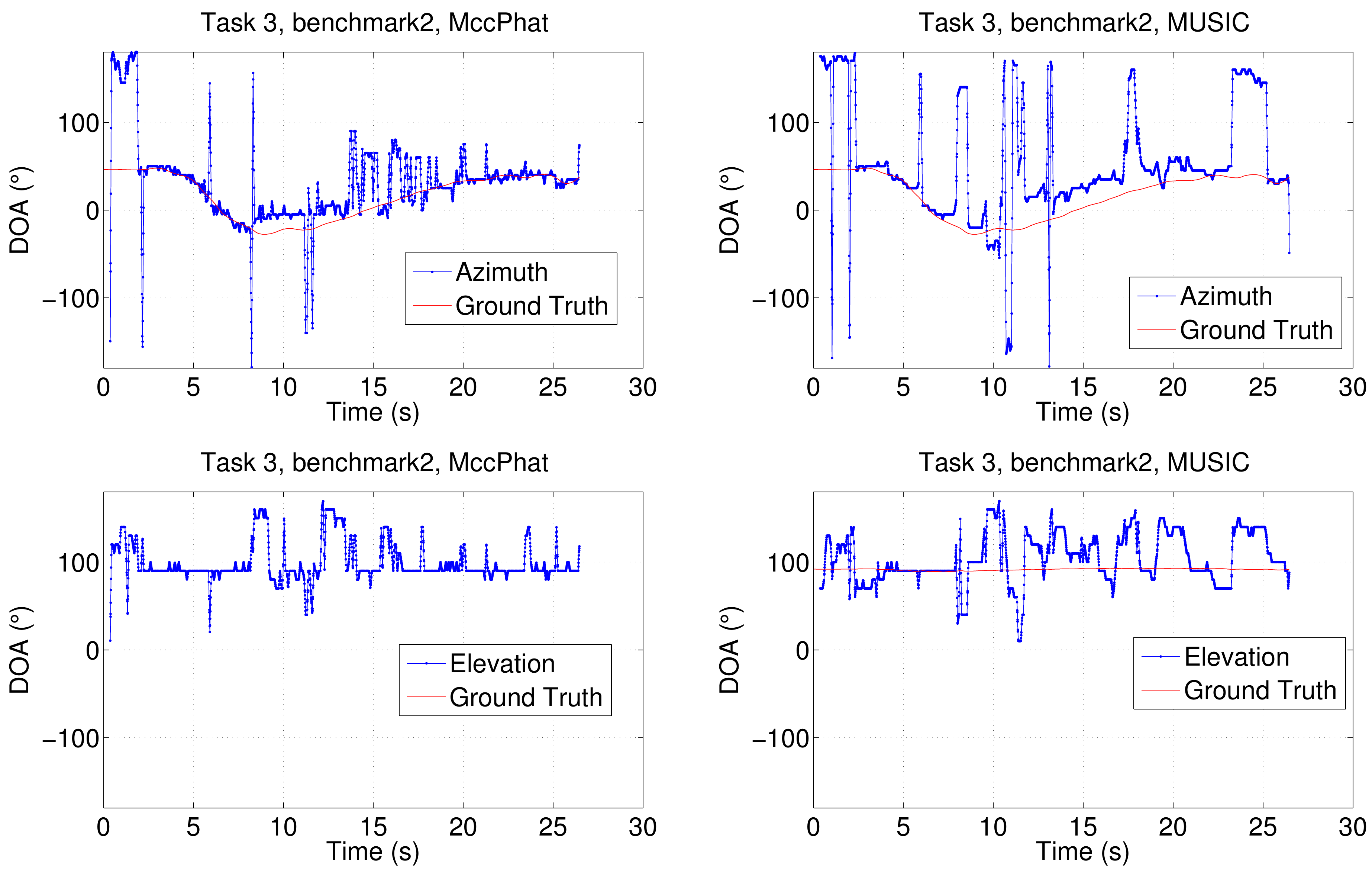}}
  \caption{Test results using benchmark2 microphone array.}
  \label{fig:Task3benchmark}
\end{figure}

\subsubsection{DICIT}

Fig.~\ref{fig:Task3dicit} shows the test results using the dicit microphone array. 
\begin{figure}[!h]
  \centering
  \centerline{\includegraphics[width=\columnwidth]{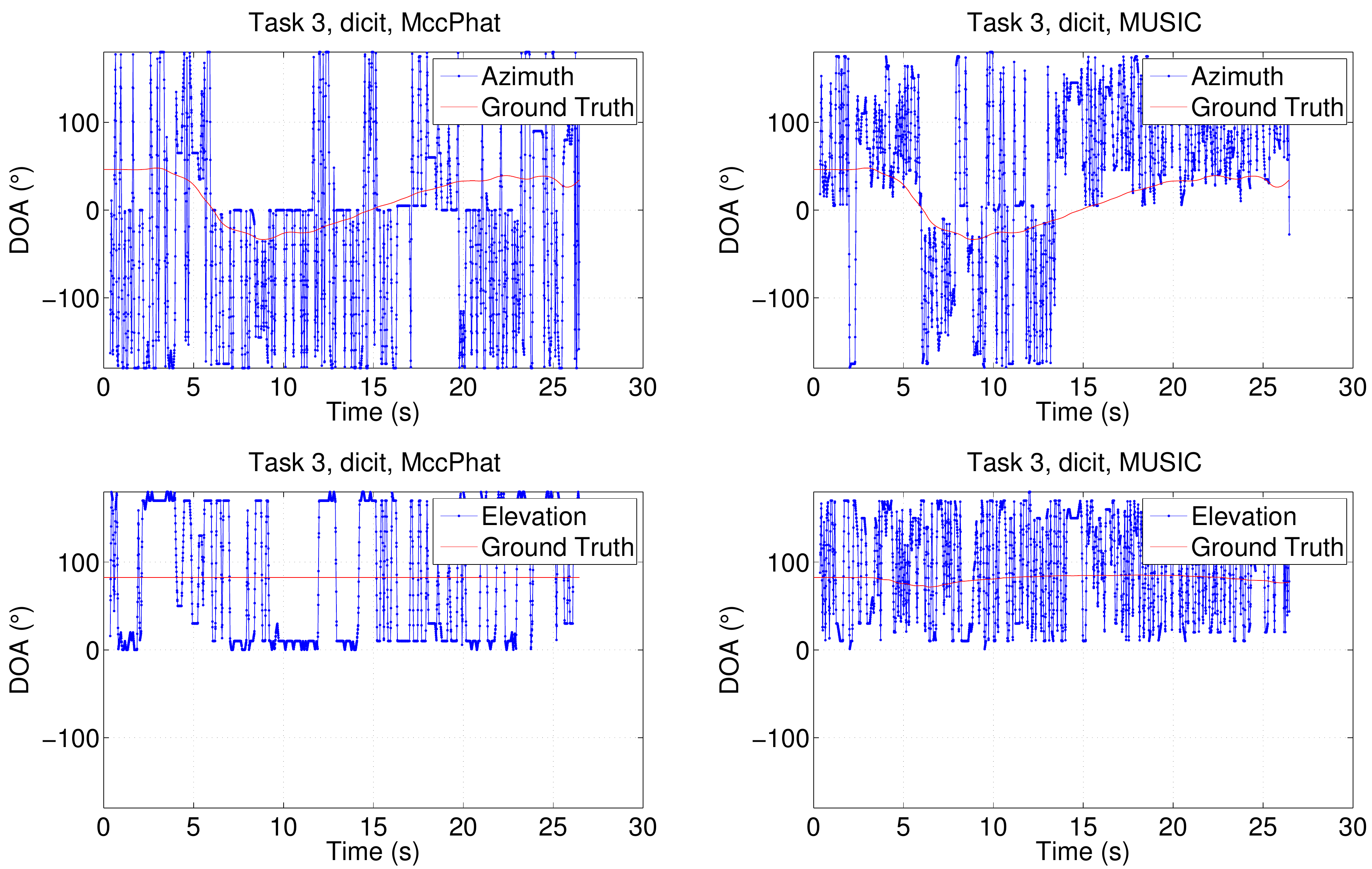}}
  \caption{Test results using dicit microphone array.}
  \label{fig:Task3dicit}
\end{figure}

\subsubsection{Dummy}

Fig.~\ref{fig:Task3dummy} shows the test results using the dummy microphone array. 
\begin{figure}[!ht]
  \centering
  \centerline{\includegraphics[width=\columnwidth]{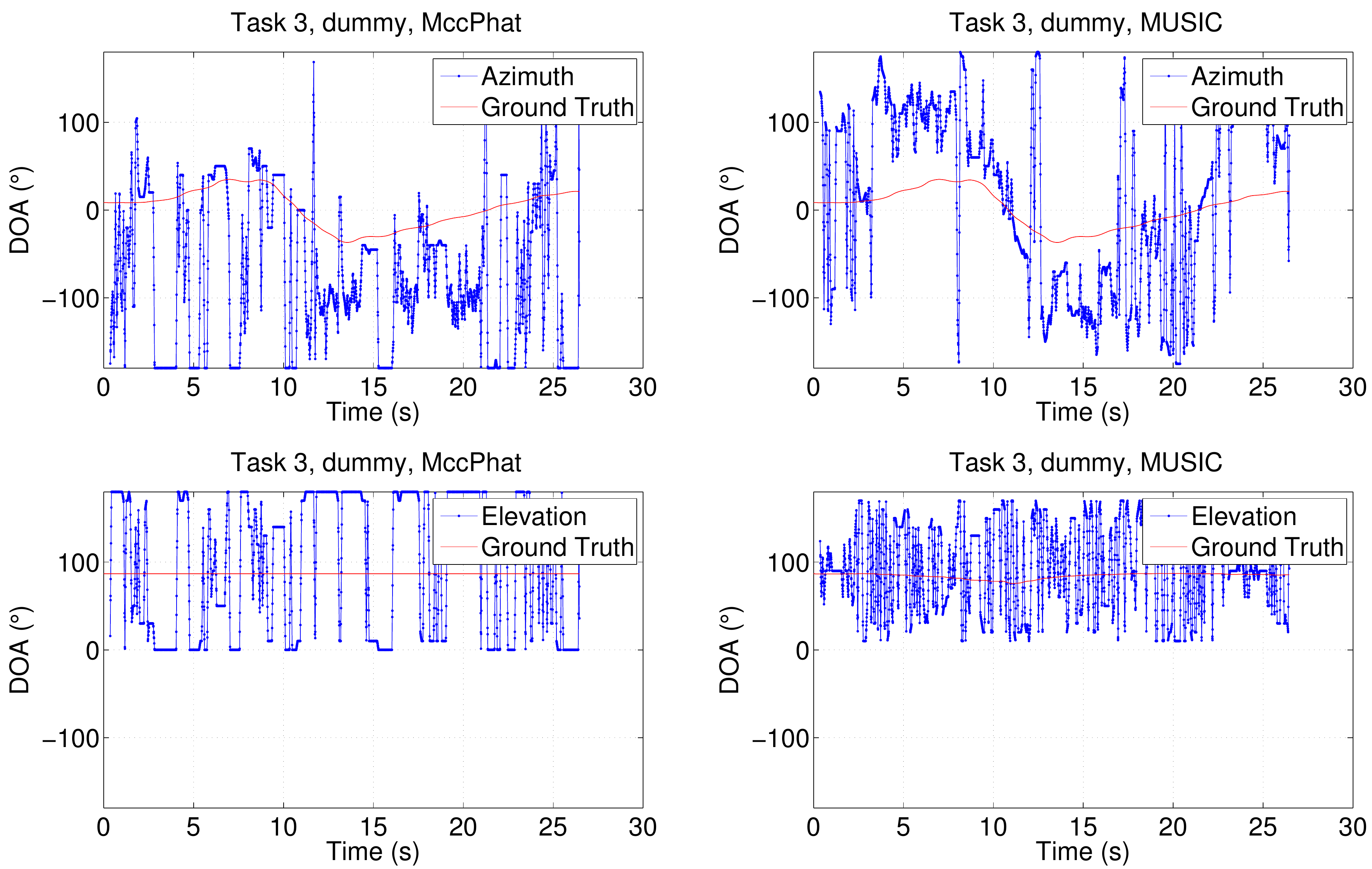}}
  \caption{Test results using dummy microphone array.}
  \label{fig:Task3dummy}
\end{figure}

From Fig.~\ref{fig:Task3benchmark} to~\ref{fig:Task3dummy}, both localization methods cannot work reliably for the moving speaker using the dicit and dummy microphone arrays. Using the benchmark2 microphone array however, can still provide locations close to ground truth. 


\subsection{Accuracy Measures}
\label{sec:ospa}

\begin{figure*}[!ht]
  \centering
  \centerline{\includegraphics[width=\textwidth,height=5.5cm]{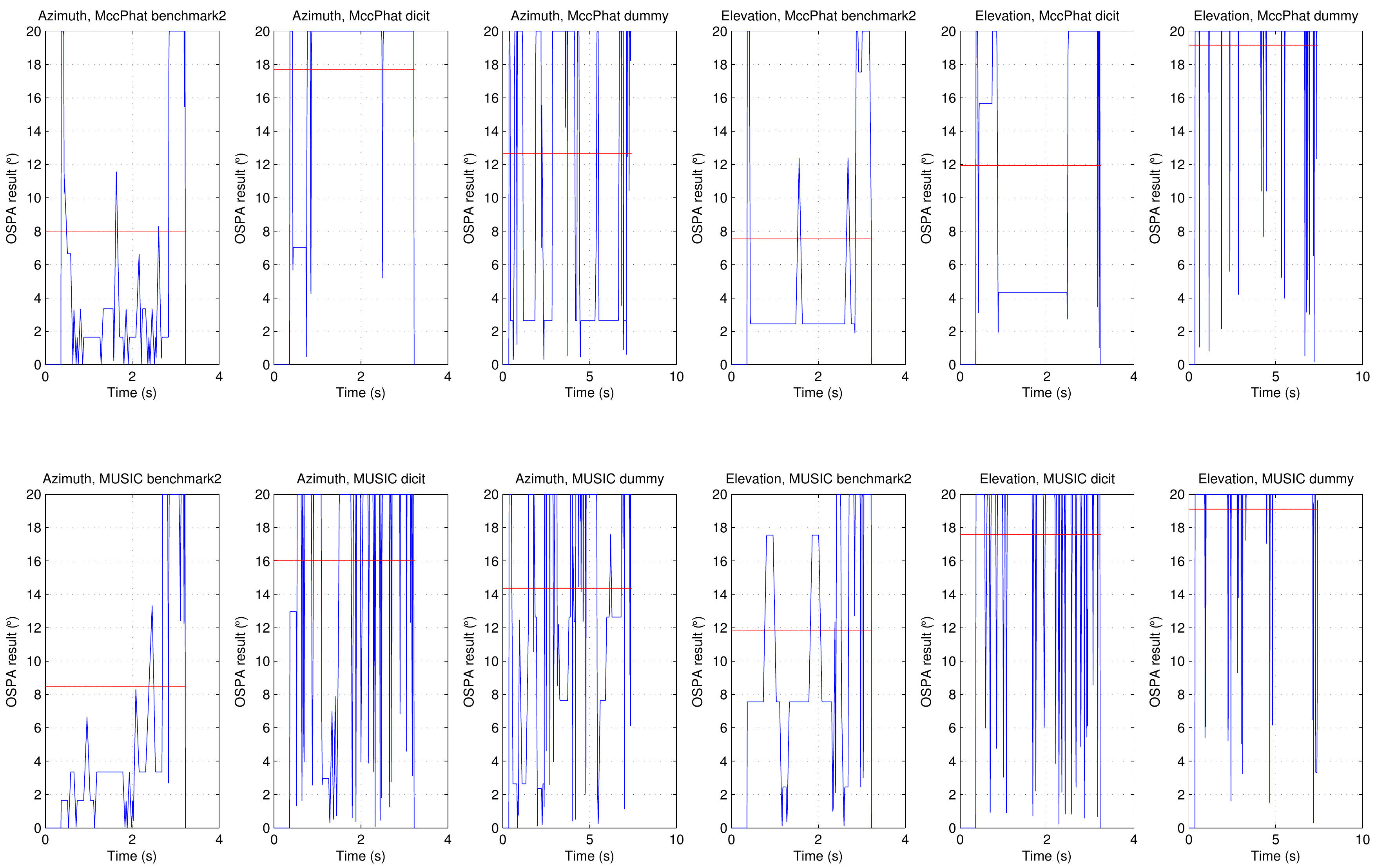}}
  \caption{OSPA results for Task1. The top row is for the MCC-PHAT method, and the bottom row is for the MUSIC method. The left three columns are for azimuth angles, while the right three columns are for elevation angles. The RMSEs are plotted in red lines. }
  \label{fig:ospaTask1}
\end{figure*}
As discussed in \cite{lin2018reverb}, the OSPA metric \cite{schuhmacher2008consistent} is used for quantitative performance measure for localization methods, when there are cardinality errors. 
When no cardinality error is taken into consideration, the OSPA metric can be simplified to the root-mean square error (RMSE) metric, by choosing the power parameter to $2$. Note however that here the cut-off parameter is chosen as $20^\circ$ for the OSPA metric. 
\begin{figure}[!ht]
  \centering
  \centerline{\includegraphics[width=\columnwidth]{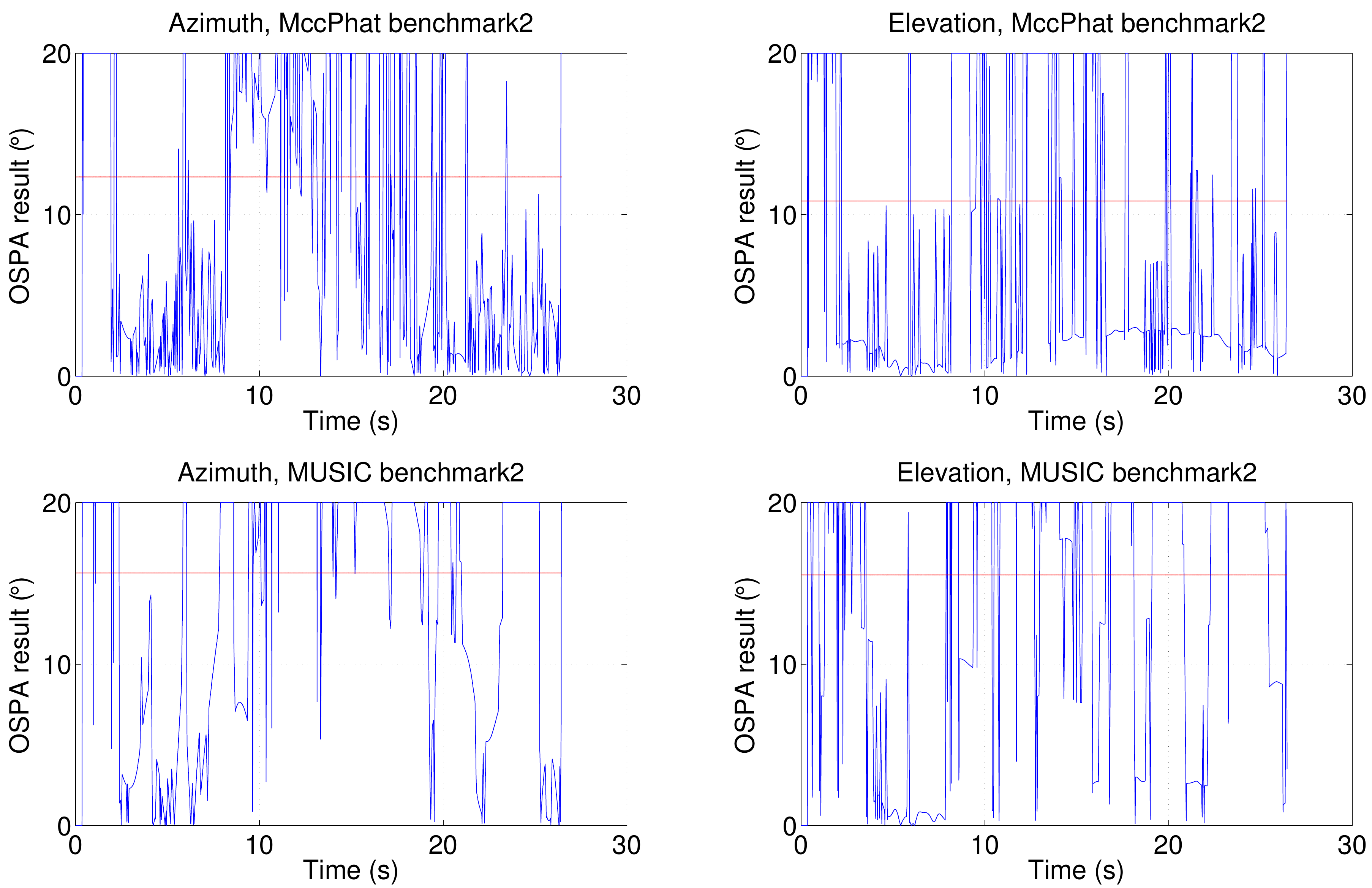}}
  \caption{OSPA results for Task3. The top row is for the MCC-PHAT method, and the bottom row is for the MUSIC method. The left column is for azimuth angles, while the right column is for elevation angles. The RMSE is plotted in a red line. }
  \label{fig:ospaTask3}
\end{figure}
Fig.~\ref{fig:ospaTask1} and Fig.~\ref{fig:ospaTask3} provide the OSPA results for the localization methods in Task1 and Task3 respectively. 
It can be seen that, except the dicit microphone for Task1, the MCC-PHAT is in general more accurate than the MUSIC method in the test cases. 

\subsection{Run Time Comparison}

The run times of respective localization methods for Task1 and Task3 and used recordings are given in Table~\ref{tab:runtime1} and Table \ref{tab:runtime2}.
Matlab implementations of algorithms are used, and the computer hardware configuration is Intel i5-3570 3.4GHz, with 32GB RAM.

\begin{table} [h]
\caption{Run time (s) for Task1 and respective recordings} \label{tab:runtime1}
\centering
\begin{tabular}{| l | l | l | l |}
    \hline
	 & benchmark2 & dicit & dummy \\
	\hline
	MUSIC & 23.528503 & 21.401130 & 50.084127 \\
	\hline
	MCC-PHAT & 241.831558 & 9.719651 & 47.801081 \\
    \hline
\end{tabular}
\end{table}

\begin{table} [h]
\caption{Run time (s) for Task3 and respective recordings} \label{tab:runtime2}
\centering
\begin{tabular}{| l | l | l | l |}
    \hline
	 & benchmark2 & dicit & dummy \\
	\hline
	MUSIC & 162.105479 & 157.904877 & 152.246116 \\
	\hline
	MCC-PHAT & 2083.525231 & 83.006261 & 168.907303 \\
    \hline
\end{tabular}
\end{table}

In general, the run time for respective methods depends on the computing resources and time length of recordings. 
In particular, the required time for the MCC-PHAT method depends on the number of closely placed microphone pairs, while that of the MUSIC method is more consistent (see Table~\ref{tab:runtime2}). 
However, it is clear that the MCC-PHAT method provides more accurate location estimates than the MUSIC method. 
Moreover, the localization results can be further processed with tracking filters (e.g. the GLMB filter \cite{vo2014labeled} \cite{lin2018joint}) using the adaptive birth model \cite{lin2016measurement}, although it is beyond the scope of this paper. 

\section{Discussions and Conclusions}
\label{sec:con}

This report presents the localization test results of the MCC-PHAT method, using sound recordings from various microphone arrays.
It is interesting to note that for both tasks (Task1 and Taks3), the benchmark2 microphone array provides most useful localization results, regardless of the algorithms used. Besides the algorithm, the array geometry is also critical to localization performance. 

Comparing with the classic MUSIC method, the MCC-PHAT shows superior localization accuracy, although at the cost of higher computational load when the number of closely located microphones is large. As shown in \cite{lin2018joint, lin2018reverb}, the accuracy advantage of MCC-PHAT can be more obvious in highly reverberant environments.

The reverberation time of the LOCATA corpus is unknown (it is possible that $T_{60} \leq 0.5$). However, the corpus provides interesting test data for evaluating location estimators for various scenarios. It will be even more useful to have the reverberation time provided for respective sound recordings.

%
%

%

%

\end{sloppy}
\end{document}